\documentclass[twocolumn,pre,floats,aps,amsmath,amssymb,nofootinbib]{revtex4-1}
\usepackage{graphicx}
\usepackage{bm}
\usepackage{physics}

\begin{document}

\title{Quantum Mechanics May Need Consciousness}
\author{Andrew Knight}
\affiliation{aknight@alum.mit.edu}
\date{\today}

\begin{abstract}
The assertion by Yu and Nikoli\'c that the delayed choice quantum eraser experiment of Kim \textit{et al.} empirically falsifies the consciousness-causes-collapse hypothesis of quantum mechanics is based on the unfounded and false assumption that the failure of a quantum wave function to collapse implies the appearance of a visible interference pattern.
\end{abstract}

\maketitle

In 2011, \textit{Annalen der Physik} published ``Quantum Mechanics Needs No Consciousness," in which Yu and Nikoli\'c \cite{Yu} attempt to empirically put to rest what they called the ``bizarre bridge between the mental and the physical."  To these authors and other physicists to whom academic discussion of a relationship between consciousness and quantum mechanics is regarded as unfashionable, the hypothesis that consciousness causes collapse (``CCC") of the quantum mechanical wave function is perhaps the most inconvenient and least liked of the various currently unfalsified hypotheses to explain the so-called measurement problem\footnote{Specifically, the inconsistency between these two statements: the quantum state of a system evolves linearly; and observations always yield outcomes (i.e., eigenstates of the measurement operator).} of quantum mechanics.  Despite assertions by \cite{Yu} to the contrary, the CCC hypothesis remains unfalsified.

Yu and Nikoli\'c \cite{Yu} address the ``delayed choice quantum eraser" experiment suggested by Scully and Dr\"uhl \cite{Scully} and performed by Kim \textit{et al.} \cite{Kim}, which will be described with reference to Figs.\ 1-3.  A laser beam is incident on a double slit, each slit containing a crystal that converts a laser photon, via the process of spontaneous parametric down conversion, to an entangled two-photon state of orthogonally polarized ``signal" and ``idler" photons, so named because the experiment is designed to detect each idler photon after (i.e., within the light cone of) detection of its entangled signal photon.  The laser beam produces photons that are monochromatic and spatially coherent -- thus indistinguishable -- over the width of the two slits.  A first detector $D_0$, which is placed in the far field by use of a converging lens (not shown), allows for detection of signal photons as a function of lateral displacement.  A prism directs idler photons through an optional beam splitter (``BS") to detectors $D_1$ and $D_2$, also placed in the far field.  

In Fig.\ 1, shown without the beam splitter, the prism and detectors are configured so that idler photons originating from slit A are detected by $D_1$ (and not $D_2$) while idler photons originating from slit B are detected by $D_2$ (and not $D_1$), so that detector $D_1$ is correlated with slit A and detector $D_2$ is correlated with slit B.  Without the beam splitter, it is asserted by both \cite{Yu} and \cite{Kim} that ``which-path" information is preserved.\footnote{I disagree that ``which-path" information ever exists in any of the proposed experimental set-ups.  Because the initial photons are spatially coherent over the width of the slits, all three detectors are placed in the far field, and the experimental setup prevents differentiation of slits A and B at the time of each biphoton's creation in the slits, no information exists -- or can ever exist -- to distinguish the creation of entangled pairs in either slit A or slit B.  \cite{Yu} and \cite{Kim} make the same mistake as Afshar \textit{et al.} \cite{Afshar} in assuming that the correlation of detector $D_1$ to slit A when slit B is closed and correlation of detector $D_2$ to slit B when slit A is closed imply the same correlations when both slits are open.  However, with both slits open, and with detectors $D_1$ and $D_2$ located in the far field, the wave functions emanating from slits A and B have already fully interfered (and rendered moot any ``which-path" information) long before detection.  A future measurement does not retroactively cause collapse or decoherence.  (See, e.g., \cite{Kaloyerou}.)  Nevertheless, for the sake of argument, I will describe the experiments consistently with analysis by \cite{Yu} and \cite{Kim} as if, in the absence of the beam splitter, detectors $D_1$ and $D_2$ indeed measure which-path information when both slits are open.  The opposing opinion that no which-path information ever exists only strengthens this paper's conclusions.}  If so, it would not be surprising, as confirmed by \cite{Kim}, that the distribution recorded at $D_0$ is the sum of two closely-spaced single-slit Fraunhofer distributions.  In other words, the detection of which-path information by detectors $D_1$ and $D_2$ guarantees no interference distribution at $D_0$.  

\begin{figure}[ht]
\includegraphics[width=3.2 in]{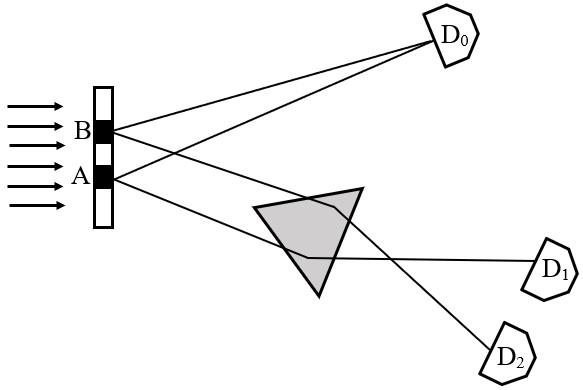}
\caption{When which-path information of idler photons is recorded by detectors $D_1$ and $D_2$, detector $D_0$ does not produce an interference pattern.}
\end{figure}

In Fig.\ 2, the beam splitter allows some photons to pass but reflects others, according to quantum mechanical randomness, so that a detection at $D_1$ is no longer correlated with emission of a photon at slit A, nor detection at $D_2$ correlated to slit B.  The beam splitter thus ``erases" any which-path information so that it is never again available, even in principle.  In a standard Young's double-slit interference experiment, the lack of which-path information is ordinarily manifested, over many photon detections, in the form of a visible interference pattern.  However, in the present case, entanglement with the idler photon complicates matters: reflection (or not) from the beam splitter produces a phase shift \cite{Kim} between the distributions correlated respectively to $D_1$ and $D_2$ so that their sum, as recorded at $D_0$, is identical to the ``no interference" distribution produced by the configuration of Fig.\ 1.  

\begin{figure}[ht]
\includegraphics[width=3.2 in]{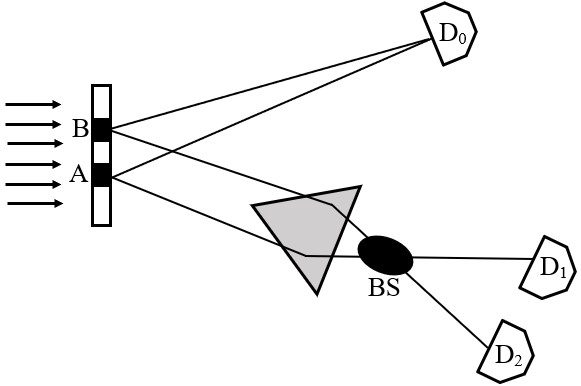}
\caption{When which-path information is erased by a beam splitter, detector $D_0$ produces a distribution that cannot be distinguished from that produced by detector $D_0$ of Fig.\ 1.}
\end{figure}

Finally, in Fig.\ 3, a coincidence counter (``CC") is placed between the detectors that allows determination of which measurements on $D_0$ correlate to detections by either $D_1$ or $D_2$.  When the distribution produced by $D_0$ corresponding only to simultaneous $D_1$ detections is plotted, a typical two-slit interference distribution appears \cite{Kim}.  The same is true of a distribution produced by $D_0$ corresponding only to simultaneous $D_2$ detections, but it is phase shifted so that the sum of these two interference patterns is indistinguishable from the ``no interference" distribution produced by the configuration of Fig.\ 1.  The analysis of \cite{Kim} suggests not only that which-path information can be forever erased from the universe, but also that this erasure can, through the use of coincidence counting and post-measurement analysis, be confirmed by the appearance of a visible interference pattern.

\begin{figure}[ht]
\includegraphics[width=3.2 in]{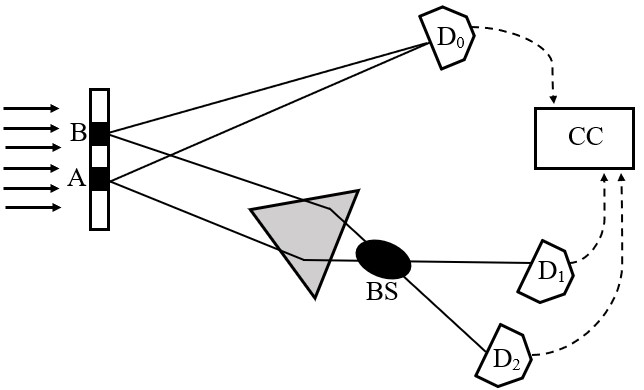}
\caption{When a coincidence counter is used post-measurement to correlate signal and idler photons, detector $D_0$ produces a visible interference pattern for signal photons that correlate to idler photons detected by detector $D_1$.}
\end{figure}

Ref.\ \cite{Yu} aims to interpret the results of \cite{Kim} as providing empirical evidence to falsify the CCC hypothesis.  The paper first characterizes the CCC hypothesis as $(\neg PR\implies \neg CWF)$, where CWF means ``collapse of the wave function" and PR means ``phenomenal representation," or the ``registering [of] the results of a measurement in consciousness," or simply conscious observation.  To falsify this statement, all that is needed is an example in which there is a wave function collapse without a corresponding conscious observation.  The paper provides no such example, and that's not surprising.  How does one provide an example of a wave function collapse without making a conscious observation?  How does one definitely state, ``There's been a wave function collapse, but it's not correlated to any conscious observation"?  What (presumably conscious) observer could possibly say that?

This conundrum has always been a problem with falsifying the CCC hypothesis.  It is not trivial.  There are, in fact, proposals for doing so, such as Deutsch's creative attempt \cite{Deutsch} to show how to empirically distinguish whether or not Wigner's Friend, presumed conscious, collapses the quantum wave function.  Proposals like this, however, heavily depend on whether there is any natural limit to the size of objects subject to interference experiments, to what extent decoherence prevents macroscopic quantum superpositions \cite{Haroche}, whether conscious perception is related to irreversible processes or the cosmological arrow of time \cite{Aaronson, Maccone}, and whether a conscious being could survive the thermal isolation necessary for a relevant interference experiment \cite{Barros}\footnote{Indeed, \cite{Barros} correctly identifies a major flaw in \cite{Yu} by pointing out why the phase shift caused by the beam splitter would, without coincidence counting, result in an apparent lack of any interference pattern detected by $D_0$, whether or not which-path information existed.}.  These are complicated and difficult questions that transcend the analysis of \cite{Yu}.

Instead of providing evidence for a wave function collapse without conscious observation, \cite{Yu} later asserts that the CCC hypothesis is equivalent to the statement, ``The interference pattern should be visible if ‘which-path' information has not been registered in consciousness of the observer," which I'll label as $(\neg PR\implies VIP)$ (``visible interference pattern").  This is a big jump from $(\neg PR\implies \neg CWF)$, and only follows, logically, if $(\neg CWF\implies VIP)$, but the paper more or less states this anyway.  (``If the photons are always in a superposition state, after a sufficient number of photons have been registered at $D_0$, [they] will exhibit the standard Young's double-slit interference pattern," and  ``Thus, the presence of the interference pattern at $D_0$ indicates whether the wave function of the original photon [sic] collapsed or not.")  In other words, \cite{Yu} positively asserts, through statements as well as logical necessity, that $(\neg CWF\implies VIP)$.  Based on their new characterization of the CCC hypothesis -- i.e., $(\neg PR\implies VIP)$ -- all the authors must do to falsify it is to give an example in which there is not a visible interference pattern when there is no conscious observation.  This is easy: they then (correctly) point out that in the Kim \textit{et al.} experiment, there is ``no interference pattern in $D_0$... irrespective of what happens with the idler photons." 

The problem is not with their evidence; it is with their logic.  They are correct that an interference pattern will not be found at $D_0$, but this says nothing to disparage the CCC hypothesis, as the logical flaw is found in their assertion that $(\neg CWF\implies VIP)$.   Regarding CWF, they make it clear that ``the relevant information" is ``which path (i.e., L or R) the photons took."  Therefore, \cite{Yu} necessarily claims that if there is no collapse of the wave function, due to a lack of which-path information, then an interference pattern will be visible at $D_0$.\footnote{We can all agree that if there \textit{is} a collapse, then an interference pattern will \textit{not} be visible.  That is, $(CWF\implies \neg VIP)$.}  So how long need we wait for a collapse?  When can we officially declare that there has been no collapse and it's time to look for interference?  Note that we are discussing \textit{entangled} photons that are notoriously contemptuous of the demands of special relativity.

The authors clarify that interference will be exhibited by photons that ``are \textit{always} in a superposition state."  (Emphasis added.)  But instead of waiting until the end of eternity to verify their theory, let's choose some firm date in the future at which we will eliminate any possibility of which-path information -- in other words, let's use the actual Kim \textit{et al.} quantum eraser experiment that \cite{Yu} cites.  Let's choose some definitive time in the future after which we no longer have to worry about the possibility of collapse.  Referring back to Fig.\ 3, the detection of the idler photon is delayed beyond detection of the signal photon by 8ns, which is much longer than the 1ns response time of the detectors \cite{Kim}.  Thus, after 8ns following detection of the signal photon by detector $D_0$, we are guaranteed that any information regarding ``which path (i.e., L or R) the photons took" is forever erased and inaccessible.  And if \cite{Yu} is correct that $(\neg CWF\implies VIP)$, we should see an interference pattern at $D_0$.

But we don't.  Nor would we expect to.\footnote{An interference pattern can be inferred only by looking at joint detection rates between $D_0$ and $D_1$ (or $D_2$), and that's only possible at a point in spacetime after (i.e., within both light cones of) the occurrence of both detections.}  And a single sentence in Footnote 1\footnote{``So proper separation of sub-populations of registered photons may be needed."  These ``sub-populations" can only be separated after post-measurement correlations are accounted for.} in \cite{Yu} indicates that the authors already knew that. 

For the sake of argument, let's give \cite{Yu} the benefit of the doubt.  Let's assume the veracity of the assertion in \cite{Yu} that $(\neg CWF\implies VIP)$ and redo the Kim \textit{et al.} experiment.  Of course, there's no reason we can't delay the idler photon by a few minutes, a few hours, perhaps even a few years before the two possible photon trajectories are recombined in a quantum eraser.  How can we be sure that the idler photon isn't intercepted and detected by some conscious person during that time?  According to \cite{Yu}, we need only look for ``the presence of [an] interference pattern at $D_0$" and that will tell us ``whether the wave function of the original photon collapsed or not."  If that were true, then we could foretell whether or not the wave function of the idler photon will collapse sometime in the future by reading the output of $D_0$ today -- a problem of backward causality.  Further, let's redesign the experiment so that I can \textit{choose} whether or not to insert the beam splitter of Fig.\ 2 into the experiment any time before detection by $D_1$ or $D_2$.  I'll run the experiment today and then decide to insert the beam splitter only if the Red Sox win the World Series in 2021.  All I'll need to do is look at $D_0$ today and I'll have my answer.  Clearly, the problem of backward causality (or ``[violation of] the no-signaling condition in quantum mechanics" \cite{Barros}) invalidates the assumption that $(\neg CWF\implies VIP)$ and is fatal to the analysis of \cite{Yu}.  

It is my opinion that consciousness exists in the physical world and is therefore fair play in the field of physics, including the extent to which relativity and quantum mechanics may relate to consciousness.  The problem of consciousness may be one of the last big questions in science, and the laws of physics may very well be implicated in its successful explanation.  The authors of \cite{Yu} reject the role of consciousness in quantum mechanics and note that ``the hypothesis that consciousness causes (or at least correlates with) the collapse of the wave function" is ``not preferred by most physicists."  That may or may not be true, but appeals to consensus in any scientific paper should be a red flag.  After all, general relativity was not preferred by most physicists in 1915.

More importantly, despite what may or may not be currently fashionable to discuss in the physics academy, \cite{Yu} fails to falsify, empirically or otherwise, the CCC hypothesis.  Whether or not quantum mechanics needs consciousness, or vice versa, is yet to be known.

\end{document}